\documentclass[wcp]{jmlr}

\usepackage[utf8]{inputenc}    
\usepackage{url}            
\usepackage{booktabs}  
\usepackage{nicefrac}       
\usepackage{microtype} 

\RequirePackage{algorithm}

\usepackage{amsmath}
\usepackage{amssymb}
\usepackage{multirow}
\usepackage{dsfont}
\usepackage{xcolor,colortbl}
\usepackage{mathtools}
\usepackage{wrapfig}

\mathtoolsset{showonlyrefs=true}

\newtheorem{defi}{Definition}
\newtheorem{thm}{Theorem}

\newcommand{\argmax}{\operatorname*{argmax}}

\newcommand{\EE}{\mathbb{E}}

\newcommand{\s}{\mathcal{S}}
\newcommand{\A}{\mathcal{A}}

\newcommand{\R}{\mathcal{R}}

\newcommand{\p}{\mathcal{P}}

\newcommand{\GG}{\mathcal{G}}

\makeatletter
\def\set@curr@file#1{\def\@curr@file{#1}} 
\makeatother
\usepackage[load-configurations=version-1]{siunitx}

\jmlrvolume{xx}
\jmlryear{2020}
\jmlrworkshop{ACML 2020}

\title[]{Foolproof Cooperative Learning}

  \author{\Name{Alexis Jacq} \Email{alexisjacq@google.com}\\
  \addr Google Research, Brain Team
  \AND
  \Name{Julien Perolat} \Email{perolat@google.com}\\
  \addr DeepMind
  \AND
  \Name{Matthieu Geist} \Email{mfgeist@google.com}\\
  \addr Google Research, Brain Team
  \AND
  \Name{Olivier Pietquin} \Email{pietquin@google.com}\\
  \addr Google Research, Brain Team
 }

\jmlrvolume{129} 
\editors{Sinno Jialin Pan and Masashi Sugiyama} 

\begin{document}

\maketitle

\begin{abstract}
This paper extends the notion of learning algorithms and learning equilibriums from repeated games theory to stochastic games. We introduce Foolproof Cooperative Learning (FCL), an algorithm that converges to an equilibrium strategy that allows cooperative strategies in self-play setting while being not exploitable by selfish learners. By construction, FCL is a learning equilibrium for repeated symmetric games. We illustrate the behavior of FCL on symmetric matrix and grid games, and its robustness to selfish learners. 
\end{abstract}

\section{Introduction}

Learning equilibriums describe a set of learning algorithms used by a team of agents to learn repeated games such that no agent can modify its algorithm and increase its average payoff~\citep{brafman2003efficient}. In that setting, a learning algorithm is viewed as a strategy that requires no initial information about the played game, so it works with any game of a given set. 

Besides, just like the Folk theorem allows to construct cooperative Nash equilibriums for repeated games based on retaliation strategies, it is possible to construct a cooperative learning equilibrium. Therefore, a team discovering a game involving a social dilemma can agree on a set of learning rules that prevent an eventual selfish player to learn a defecting behavior. In this paper, we build a cooperative learning equilibrium that applies to the more general set of stochastic games~\citep{shapley1953stochastic} involving sequential decision making.

Multi-agent reinforcement learning (MARL) brings a framework to construct algorithms that aim to solve stochastic games where players individually or jointly search for an optimal decision-making policy to maximize a reward function.
Individualist approaches mostly aim at reaching Nash equilibriums, taking the best actions whatever the opponents behaviors are~\citep{bowling2001rational,littman2001friend}. Cooperative approaches aim at optimizing a common objective.
If agents can agree on joint actions, cooperation can be viewed as a single agent problem in a larger dimension~\citep{claus1998dynamics}. However, when rewards are individual but agents agree to learn a joint cooperative approach, they remain easily exploited if one agent starts being individualist. In that case, a team needs a learning equilibrium that safely leads rational agents to learn the common objective rather than the maximization of their own rewards.

We focus on symmetric situations where no agent has an individual advantage. For example, this is the case on an island with a quantity of resources equally accessible to all agents. Moreover, we consider repeated stochastic games~\citep{de2012polynomial}, modelling the situation where players restart from the beginning after reaching absorbing states or after a finite number of time steps. In the island resource example, repetitions could represent successive days or 4-seasons cycles. In fact, any repeated stochastic game where all players have the same reward function and are subject to the same transition laws is symmetric. This applies to most of common-pool resource appropriation games~\citep{perolat2017multi} and other sequential social dilemmas~\citep{leibo2017multi}. 

In this context, we extend the definition of learning algorithms -- mapping between histories and probability distributions -- and the definition of learning equilibriums to stochastic games. We introduce Foolproof Cooperative Learning (FCL), an algorithm that both learns cooperative and retaliating strategies. In that perspective, one FCL player $i$ learns two value-functions for each other player $j$, the one estimating the $j$'s payoff for cooperating when all the team cooperates, and the other estimating $j$'s best possible payoff when all the team tries to lower its score in an eventual retaliation. When a player deviates from the agreed exploring behavior and is not cooperative during greedy steps, the rest of the team retaliates in consequence. We show that FCL is a learning equilibrium forcing a cooperative behavior, and we empirically verify this claim with two-agent matrix games and grid-world repeated symmetric games. 
Our contributions are (1) a theoretical framework and an experimental setting to study learning equilibriums for stochastic games and (2) the construction of a cooperative learning equilibrium using finite retaliations and so allowing selfish learners to eventually cooperate after exploring defections.

\section{Definitions and Notations}

An $N$-player stochastic game can be written as a tuple $(\s, (\A_i)_{i=1\dots N}, \p, \mu_0, (r_i)_{i=1\dots N})$, where $\s$ is the set of states, $\A_i$ the set of actions for player $i$, $\p$ the transition probability ($\p(\cdot  \vert s, a_1\dots a_N)$), $\mu_0$ a distribution over initial states ($\mu(s^0)$), $r_i$ the reward function for player $i$ ($r_i(s, a_1\dots a_N)$). We also assume bounded, deterministic reward functions and finite state and action spaces. In a repeated stochastic game, a stochastic game (the stage game) is played and terminates when at least one player reaches an absorbing state, or after a finite number of steps. This is repeated an infinite number of times, and players have to maximize their average return during a stage game \citep{de2012polynomial}. 

A stationary strategy for player $i$, $\pi_i(\cdot \vert s)\in\Pi_{\A_i}$, maps a state to a probability distribution over its set of possible actions. We write $\pi_{-i}$ the product of all players strategies but player $i$ and $\boldsymbol{\pi} = \pi_1\times\dots\times\pi_N = \pi_i\times\pi_{-i}$ the product of all players strategies (the strategy profile). Given opponents strategies $\pi_{-i}$, the goal for a rational player $i$ is to find a strategy $\pi_i^*$ that maximizes its average payoff $\R_i$ during a stage game:
\begin{align}
    \pi_i^* &= \argmax_{\pi_i}\R_i(\pi_i, \pi_{-i}) = \argmax_{\pi_i}\EE_{\pi_i, \pi_{-i}, \p}\sum_l \gamma^l r(s^l, a_i^l, a_{-i}^l).
\end{align}
The policy $\pi_i^*$ depends on the opponents strategies and is called the best response for player~$i$ to $\pi_{-i}$. In general, we call strategy any process $\{\pi^t\}_t$ defining a stationary strategy for any stage $t$. The value of a player's non-stationary strategy $\{\pi^t\}_t$ is the average return over stage games, $\EE_{t>0}\left[\R_i(\pi^t_i, \pi^t_{-i})\right]=\liminf_{T\rightarrow\infty}\EE\left[\sum_{t=0}^T \R_i(\pi^t_i, \pi^t_{-i})\right]$. A Nash equilibrium describes a stationary strategy profile $\boldsymbol{\pi}^*=\pi_1^*\times\dots\times\pi_N^*$, such that no player can individually deviate and increase its payoff \citep{nash1951non}:
\begin{equation}
    \forall 1\leq i\leq N,\; \forall \pi_i \in\Pi_{\A_i}, \; \R_i(\pi_i, \pi_{-i}^*) \leq \R_i(\pi_i^*, \pi_{-i}^*).
\end{equation}
 This definition can be extended to non-stationary strategies using expected return over stage games: no player can individually deviate from an equilibrium non-stationary strategy and increase its average payoff:
\begin{align}
    \forall 1\leq i\leq N,\; \forall \{\pi_i^t \in\Pi_{\A_i}\}_t, \;
    \EE_{t>0}\left[\R_i(\pi_i^t, {\pi_{-i}^{t,*}})\right] \leq \EE_{t>0}\left[\R_i({\pi_i^{t,*}}, {\pi_{-i}^{t,*}})\right].
\end{align}
As $\EE_{t>0}\left[\R_i(\pi_i^t, {\pi_{-i}^t})\right]=\R_i(\pi_i, {\pi_{-i}})$ for stationary strategy profiles, any stationary strategy equilibrium is still an equilibrium among non-stationary processes.

In order to allow rewarding or retaliation strategies, we only consider games where all players are aware of all opponents actions and rewards, and receive a signal each time the game is reset. We also admit players to share information with some opponents in order to organize joint retaliation actions or joint explorations. Moreover, we only consider \textit{Repeated Symmetric Games} (RSG):
\begin{defi}[Repeated Symmetric game (RSG)]
    A repeated stochastic game is symmetric if, for any stationary strategy profile ($\pi_1\dots\pi_N$) and  any permutation $\psi$ over players:
    \begin{equation}
        \forall 1\leq i \leq N, \; \R_{\psi(i)}(\pi_i, \pi_{-i}) = \R_i(\pi_{\psi(i)}, \pi_{\psi(-i)}).
    \end{equation}
\end{defi}
This generalizes the definition for symmetric $N$-player matrix games~\citep{dasgupta1986existence} to stochastic games where players' utilities are replaced by average returns\footnote{Actually, the definition of~\cite{dasgupta1986existence}, $\forall i, \R_i(\pi_i, \pi_{-i}) = \R_{\psi(i)}(\pi_{\psi(i)}, \pi_{\psi(-i)})$, is incorrect: symmetries are not independent of player identities, which is not the case if the right-hand return is indexed with the inverse permutation instead~\citep{vester2012symmetric}.}. Note that in a symmetric game, for any Nash equilibrium with returns $(\R_1\dots\R_N)$ and for any permutation $\sigma$ over players, there is another Nash equilibrium with returns $(\R_{\sigma(1)}\dots\R_{\sigma(N)})$. In this paper, we use the concept of $N$-cyclic permutations to construct specific strategies:
\begin{defi}[$N$-cyclic permutation]
A permutation $\sigma$ is N-cyclic if for all $i,j\in\{1\dots N\}$, there is $k$ such that $\sigma^k(i)=j$.
\end{defi}

\section{Cooperation and retaliation}\label{folk}
We say a strategy is cooperative if it maximizes a common quantity $\hat{\R} = f(\R_1\dots\R_N)$. Usual examples are strategies that maximize the sum, the product or the minimum of players returns. In RSGs, the strategy that maximizes the minimum of player returns is particularly interesting as it coincides with the best egalitarian payoff. In this paper, we refer to this strategy as the \emph{egalitarian strategy}. An important property of RSGs is the fact that egalitarian strategies can always be obtained by repeatedly applying an $N$-cyclic permutation on a stationary strategy that maximizes the sum of players returns (see appendix~\ref{proof_egalitarian}):
\begin{thm}\label{egalitarian}
    Let $\pi^\Sigma_i$ be a stationary strategy for player $i$ that maximizes the sum of players returns in an N-player RSG, $\sigma$ an N-cyclic permutation over players, and $t$ indexing the successive stages. Then, the strategy $\boldsymbol{\pi}^t = (\pi^\Sigma_{\sigma^t(1)}\dots\pi^\Sigma_{\sigma^t(N)})$ (where $\sigma^t=\sigma\circ\dots\circ\sigma$ $t$ times) is an egalitarian strategy.
\end{thm}
In repeated games, players can retaliate when a selfish one deviates from a target strategy. In that case, the target strategy is said to be enforceable~\citep{osborne1994course}: if all players are accorded to retaliate when one player deviates from a strategy profile, no player can improve its payoff by individually deviating from the strategy profile. This phenomenon is described by the Folk theorem and allows building equilibrium strategies achieving cooperative goals. In an RSG, the egalitarian strategy is enforceable since for any player $i$, its payoff $V^c_i$ for cooperating is always larger than its minimax payoff $V^r_i$ when all other players retaliate to lower its reward (because of the symmetric and convex shape of the space of possible payoffs~\citep{de2012polynomial}).
When a single retaliation is too small so it is still worth defecting for a selfish player, it must be repeated. As~\cite{littman2005polynomial}, one can compute a minimal number of retaliations to enforce the cooperation. Let $V^d_i$ be the payoff obtained by player $i$ for defecting, then

\begin{equation}\label{punish_repeats}
     K_j = \left\lceil\frac{V^d_j - V^c_j}{V^c_j - V^r_j}\right\rceil
\end{equation}
is a sufficient number of retaliations (see Appendix~\ref{retaliations}). In the edge case where $V^c_j = V^r_j$, the required number of retaliations becomes infinite, but the cooperative objective is not affected (this is the case in rock–paper–scissors). Let $\boldsymbol{\pi}_\text{Folk}$ be the (non-stationary) strategy that follows $\boldsymbol{\pi}^*$ if all players cooperate, or repeat a minimax retaliation over $K_j$ stage games if a player $j$ deviates from $\boldsymbol{\pi}^*$. By construction, $\boldsymbol{\pi}_\text{Folk}$ is a Nash equilibrium.

\section{Learning algorithm}

In normal form games, a \textit{learning algorithm} maps an history of actions and rewards to a probability distribution~\citep{brafman2003efficient}. In stochastic games, one must consider a mapping between an history of transitions during stage games, $\mathcal{H}^T = \{\{s^l, a_i^l, a_{-i}^l, r_i^l, r_{-i}^l\}_{l\in t}\}_{t<T}$, to state-depending strategies $\pi(.\vert s)$.
For simplification, we will write\footnote{\cite{brafman2003efficient} directly treated learning algorithms like strategies (considering the meta game of choosing an algorithm for a set of games). In this paper, we preferred to keep a distinct notation in order to stay consistent with the reinforcement learning notation.} $A_i(T) = A_i(\mathcal{H}^T) =\pi_i^T$ the algorithm used by a player $i$.
The algorithm profile $\boldsymbol{A} = (A_1\dots A_N)$ is the set of all players algorithms.

\subsection{Multi-agent learning}
Reinforcement learning provides a class of algorithms that aim at maximizing an agent's return. Out of all of them, our interest concerns $Q$-learning approaches~\citep{watkins1992q} for three reasons: they are model-free, off-policy and they are guaranteed to converge in finite state and action spaces. In a game $\GG$, for a player $i$ and given opponents policy $\pi_{-i}$, the basic idea is to learn a $Q$-function that approximates, for each state and action, the average return starting from playing that specific  action at that given state while using the best strategy afterwards. For the $Q$-function $Q_i$, associated with player $i$'s optimal policy (the one that maximizes its return), the following holds:
\begin{align}
    Q_i(s,a_i&, a_{-i}) = r_i(s,a_i, a_{-i})+ \gamma\sum_{s'}\p(s'\vert s, a_i, a_{-i})\max_{a_i'}Z_i(a_i', s', \pi_{-i}),
\end{align}
where $Z_i(a_i, s, \pi_{-i}) = \sum_{a_{-i}'}\pi_{-i}(a_{-i}\vert s')Q_i(s', a_i', a_{-i}')$ is the expected value for agent $i$ given its opponents policy. The $Q$-learning algorithm is constructed in order to progressively approximate the $Q$-function without knowledge of the problem dynamics $\p$ and reward functions $r$, and without knowing the decision process that generated the history buffer (in contrast, for example, to policy gradient algorithms~\citep{williams1992simple}). In finite state and action spaces, the approximation is obtained by successively applying the updates:
\begin{align}\label{Qupdate}
    Q&_i^{t+1}(s^t,a_i^t, a_{-i}^t) = Q_i^t(s^t,a_i^t, a_{-i}^t)+\alpha_t \big(r_i^t + \gamma \max_{a_i} Z_i(a_i,s^{t+1}, \pi_{-i}) - Q_i^t(s^t, a_i^t, a_{-i}^t)\big),
\end{align}
where $\alpha_t$ is the learning rate. However, when the opponent policy is not fixed, maximizing the $Q$-function with respect to actions is no longer an improvement of the policy (the response of the opponents to this deterministic policy can decrease the average player's return). MARL provides several alternative greedy improvements. For example, a defensive player can expect opponents to minimize its $Q$-function (\textit{minimax} $Q$-learning). In that case, a greedy improvement of the policy to evaluate the value of a new state is obtained by solving the linear problem \citep{littman1994markov}:
\begin{equation}\label{minimax}
    \pi_i^{\textit{greedy}}(.|s) = \argmax_{\pi_i}\min_{a_{-i}}\sum_{a_i}\pi_i(a_i\vert s)Q_i(s, a_i, a_{-i})
    = \argmax_{\pi_i}\min_{a_{-i}} Z_{-i}(a_{-i}, s, \pi_i)
\end{equation}
and the corresponding $Q$-learning update becomes:
\begin{align}\label{minimaxupdate}
    &Q_i^{t+1}(s^t,a_i^t, a_{-i}^t) = Q_i^t(s^t,a_i^t, a_{-i}^t)+\alpha_t \big(r_i^t + \gamma \max_{\pi_i}\min_{a_{-i}}Z_{-i}(a_{-i}, s, \pi_i) - Q_i^t(s^t, a_i^t, a_{-i}^t)\big).
\end{align}
Besides, even when the game is cooperative, the fact that all strategies are changed at each update breaks the Markov hypothesis. If agents cannot communicate, it becomes complex to understand how one agent personally contributed to improving or deteriorating the strategy profile~\citep{foerster2018counterfactual}. This is not the case in this paper, as we consider the case where agents can fully communicate. Thus, learning a cooperative joint strategy becomes equivalent to solving a larger dimensional MDP and the optimal (cooperative) strategy is iteratively computed:
\begin{equation}\label{cooperupdate}
    Q^{c,t+1}(s^t,a_i^t, a_{-i}^t) = Q^{c,t}(s^t,a_i^t, a_{-i}^t)+
    \alpha_t \big(\sum_i r_i^t + \gamma \max Q^{c,t}(s^{t+1},.)- Q^{c,t}(s^t, a_i^t, a_{-i}^t)\big).
\end{equation}
In order to learn a retaliating behavior with a finite number of retaliations, a team of agents must learn a defensive strategies against a single defecting agent ($j$), which corresponds to the opposite of \textit{minimax} $Q$-learning indexing where one agent is defensive against others:
\begin{equation}\label{retaliatupdate}
    Q_j^{r,t+1}(s^t,a_i^t, a_{-i}^t) = Q_j^{r,t}(s^t,a_i^t, a_{-i}^t)+
    \alpha_t \big(r_i^t + \gamma \max_{a_{j}}\min_{\pi_{-j}}Z_{j}(a_{j}, s, \pi_{-j}) - Q_j^{r,t}(s^t, a_i^t, a_{-i}^t)\big).
\end{equation}
For estimating a finite number of retaliations with Eq.~\eqref{punish_repeats}, agents must learn the maximum benefit one agent can obtain by defecting while others are cooperating, $V_j^d$. Knowing $Q^c$, this value can also be derived from Q-learning:
\begin{align}\label{defectupdate}
    Q_j^{d,t+1}(s^t,a_i^t, a_{-i}^t) &= Q_j^{d,t}(s^t,a_i^t, a_{-i}^t)+\alpha_t \left(\sum_i r_i^t + \gamma V_j^d(s^{t+1})- Q^{d,t}(s^t, a_i^t, a_{-i}^t)\right)\\
    \text{with }V_j^d(s) &= \max_{a_{j}'} Q^d_j(s, a_j', \argmax_{a_{-j}}\max_{a_j}Q^c(s, a_j, a_{-j})).
\end{align}

\subsection{Learning equilibrium}

Learning equilibriums are defined over normal form games~\citep{brafman2003efficient}, but one could consider any stochastic game as a normal form game where possible decisions are the set of all possible strategies $\pi(.|s)\in\Pi_\A$. In that sense, the original definition applies to stochastic games: 
\begin{defi}[Learning equilibrium]
    Let $\GG$ be a set of stochastic games. An algorithm profile $\boldsymbol{A}^* = (A_1^*\dots A_N^*)$ is a learning equilibrium for $\GG$ if, for any game $g \in\GG$, there is a stage $T_g$ such that, for any player $i$ and any learning algorithm $A_i$:
    \begin{align}
       \EE_{t>T_g}\R_i\big(A_i(t), A_{-i}^*(t)\big) \leq \EE_{t>T_g} \R_i\big(A_i^*(t), A_{-i}^*(t)\big).
    \end{align}
\end{defi}
Just like for Nash equilibriums, no player can individually follow an alternative algorithm and increase its asymptotic score. However, one important difference is the fact that a learning algorithm is not defined with respect to a particular game, but a set of games. We may think that a process always playing a Nash equilibrium of the given game ($\pi_i^t = \pi_i^*$ for all $t$) is a learning equilibrium. However, such a process requires an initial knowledge of the game and can't be obtained from a process starting with an empty condition. Therefore, it can't be described as a learning algorithm. For the same reason, the strategy $\boldsymbol{\pi}_\text{Folk}$ described in Sec.~\ref{folk} is not a learning equilibrium. But one can construct a learning equilibrium that learns to play $\boldsymbol{\pi}_\text{Folk}$: this is the idea of FCL.

\section{Foolproof cooperative learning}

\begin{algorithm}[h]
    \caption{FCL for player $i$.}
    \label{algo:FCL}
    
\textbf{Input}: initial functions $Q^c$, $\{Q^d_i\}_{i=1\dots N}$ and $\{Q^r_i\}_{i=1\dots N}$, and N-cyclic permutation $\sigma$.

\textbf{For} stages $t=1$ \textbf{to} $+\infty$:

    \hspace{5mm} \textbf{While} stage continue
    
        \hspace{10mm} \textbf{If} no need for retaliation
        
            \hspace{15mm} If exploration time: \textbf{explore} with $a_i \sim \textit{uniform}$
                
            \hspace{15mm} Otherwise: \textbf{cooperate} with $a_i \sim \pi^{c}_{\sigma^t(i)}(Q^c)$
                
        \hspace{10mm} \textbf{Else} (player $j$ previously defected)
        
            \hspace{15mm} \textbf{retaliate} with $a_i \sim \pi^{r}_{i,j}(Q^r_j)$
            
        \hspace{10mm}Observe $a_{-i}$ and new state $s'$, receive reward $r_i$ and observe $r_{-i}$
        
        \hspace{10mm}Update $Q^c$ according to~\eqref{cooperupdate}
        
        \hspace{10mm} \textbf{For} all other agents $j\neq i$
        
            \hspace{15mm} Update $Q^r_j$ according to~\eqref{retaliatupdate}
            
            \hspace{15mm} Update $Q^d_j$ according to~\eqref{defectupdate}
            
            \hspace{15mm} \textbf{If} it was not an exploration and $j$ did not cooperate
            
                \hspace{20mm} Compute the number of required retaliations according to~\eqref{punish_repeats}
\end{algorithm}

FCL, as described in Alg.~\ref{algo:FCL} (for a player $i$), converges to $\boldsymbol{\pi}_\text{Folk}$ over the explored set of states. In an $N$-player game, each FCL player approximates $2N+1$ $Q$-functions: one associated with the cooperative policy that maximizes the sum of all players ($Q^c$), $N$ associated with retaliation policies preventing any defection from other players $j$ ($Q^r_j$), and $N$ associated with each opponent's best response to the cooperative strategy (learned through $Q^d_j$ with Eq.\eqref{defectupdate}). At each stage game, FCL will play according to an egalitarian strategy (learned through $Q^c$ with Eq.\eqref{cooperupdate} and applying an N-cyclic permutations to the associated strategy profile) unless one of the opponents deviates from that strategy. In case of a defection, all FCL agents will agree on a joint retaliation based on a defensive strategy (learned through $Q^r_j$ with Eq.\eqref{retaliatupdate}) for $K$ stages (according to Eq.\eqref{punish_repeats}). In order to allow exploration, a deterministic process is used to decide, at each time $t$, between exploration and exploitation. We highlight that FCL differs from pure \textit{minimax} Q-learning in the fact that it also learns to cooperate, and differs from Friend-or-Foe $Q$-learning (FFQ) \citep{littman2001friend} which learns to play cooperatively if an opponent is cooperative, or defensively if the opponent is defective with a single $Q$-function, while FCL learns both cooperation and defense in a disentangled way. 
Thm~\ref{fcl} states that FCL is a learning equilibrium for RSGs forcing a cooperative behavior.
\begin{thm}\label{fcl}
    FCL is a learning equilibrium for RSGs.
\end{thm}
In the proof of Thm~\ref{fcl} (see Appx.~\ref{proof_fcl}), we show that any team of communicating FCL agents converges to $\boldsymbol{\pi}_\text{Folk}$. As constructed, FCL can be seen as a team rule designed to prevent the deviation of one agent of the team. This assumes that agents can be centralized (in order to communicate the times of exploration) and that no sub-team of agent can simultaneously start to deviate and defect. In the case where agents are unable to communicate, they need to estimate if a deviation is caused by an exploration or is resulted by the exploitation of a selfish behaviour. Such an ability, studied in the context of simple matrix games~\citep{ashlagi2012robust}, would likely improve the robustness of FCL which, so far, requires a relatively high level of coordination. If more than one agent simultaneously start to defect, this sub-team could be perceived as one 
``meta-opponent'' to punish during a longer time. This opens another perspective of future improvement. 

\section{Experiments}

\begin{table}[ht]
    \centering
    \caption{Payoff matrices used for IPD, aIPD, ICH and RPS. }
    \label{matrix_games}
    \begin{tabular}{ccc}
        IPD & aIPD\\ 
        \begin{tabular}{|c|c|c|}
            \hline 
            \cellcolor{gray!50} & C & D
            \\
            \hline 
            C & (-1,-1) & (-3,0)
            \\
            \hline
            D & (0,-3) & (-2,-2)
            \\
            \hline
        \end{tabular}
        &
        \begin{tabular}{|c|c|c|}
            \hline 
            \cellcolor{gray!50} & C & D
            \\
            \hline 
            C & (-1,-1) & (-3,10)
            \\
            \hline
            D & (10,-3) & (-2,-2)
            \\
            \hline
        \end{tabular}
        \\
        ICH & RPS\\
        \begin{tabular}{|c|c|c|}
            \hline 
            \cellcolor{gray!50} & Swerve & Straight
            \\
            \hline 
            Swerve & (2,2) & (1,3)
            \\
            \hline
            Straight & (3,1) & (0,0)
            \\
            \hline
        \end{tabular}
        &
        \begin{tabular}{|c|c|c|c|}
            \hline 
            \cellcolor{gray!50} & R & P & S
            \\
            \hline 
            R & (0,0) & (-1,1) & (1,-1)
            \\
            \hline
            p & (1,-1) & (0,0) & (-1,1)
            \\
            \hline
            S & (-1,1) & (1,-1) & (0,0)
            \\
            \hline
        \end{tabular}
    \end{tabular}

\end{table} 

\begin{table}[ht]
    \centering
    \caption{Grid games. $A$ is the starting position of one player, $B$ is the starting position of the other. At each turn, both players simultaneously select one action among going up, down, left, right or stay. When reward cells with $\$$ symbol are reached by one player, the player obtains the corresponding reward and the game is immediately reset. $\$_{A:X}$ means that only player $A$ gets the reward $X$ when reaching the cell, $\$_{X}$ means that any player gets reward $X$ when reaching the cell, and $\$_{X,Y}$ means that the player who reaches the cell gets $X$ and the other gets $Y$ (if the other player reach another rewarding cell, the rewards are summed). Two players can not be on the same cell at the same time and they can not cross each other. In case of conflict, one player reaches the cell and the other stays with probability 0.5. Grey cells are walls and are not reachable.}
    \label{grids}
    \begin{tabular}{cc}
        (a) Grid prisoners dilemma & (b) Compromise\\
        \begin{tabular}{|c|c|c|c|c|c|c|c|c|}
            \hline 
            \cellcolor{gray!50} 
            &\cellcolor{gray!50}
            &\cellcolor{gray!50}
            &\cellcolor{gray!50}
            & $\$_{100}$
            &\cellcolor{gray!50} 
            &\cellcolor{gray!50}
            &\cellcolor{gray!50}
            &\cellcolor{gray!50}
            \\
            \hline 
            $\$_{A:100}$ & & & A & & B & & & $\$_{B:100}$
            \\
            \hline
        \end{tabular}
        &
        \begin{tabular}{|c|c|c|c|c|c|c|c|c|}
            \hline 
            &\cellcolor{gray!50}
            & $\$_{B:100}$
            &\cellcolor{gray!50}
            &
            &\cellcolor{gray!50} 
            &$\$_{A:100}$
            &\cellcolor{gray!50}
            &
            \\
            \hline 
            & & A & & & & B & & 
            \\
            \hline
        \end{tabular}
        \vspace{3mm}
        \\
        (c) Coordination & (d) Temptation\\
        \begin{tabular}{|c|c|c|}
            \hline 
            $\$_{B:100}$ & & $\$_{A:100}$
            \\
            \hline
            & &
            \\
            \hline 
            A & & B
            \\
            \hline
        \end{tabular}
        &
        \begin{tabular}{|c|c|c|c|}
            \hline 
            $\$_{20,-10}$ & A & B & $\$_{20,-10}$
            \\
            \hline 
            $\$_{40,-20}$ & & & $\$_{40,-20}$
            \\
            \hline
        \end{tabular}
    \end{tabular}

\end{table}  

We first explored the case of well known repeated symmetric matrix games involving two players: Iterated Prisoners Dilemma (IPD), Iterated Chicken (ICH) and Rock-Paper-Scissors (RPS). We also studied an altruist version of the IPD (aIPD), where the optimal way to share the outcomes in average is when the two players alternate between (player 1 cooperate, player 2 defect) with rewards (-3, 10) and (player 1 defect, player 2 cooperate), with rewards (10, -3): in average, both players get 3.5. Table~\ref{matrix_games} shows the payoff matrices. Then, we investigated larger state spaces with grid games known to induce coordination problems and social dilemmas~\citep{de2012polynomial}. We introduced a new grid game, closer to the concept of limited resource appropriation: the Temptation game. In Temptation, making a movement to the sides can be seen as taking immediately the resource, while making a movement to the bottom can be seen as waiting for the winter. All grid games are described in details in Table~\ref{grids}. In order to verify that FCL is a learning equilibrium, we compare the score obtained by FCL and by selfish learning algorithms, $Q$-learning and policy-gradient (PG), against FCL. Indeed, we expect a learning equilibrium performing better than any other algorithm when the opponents are following the learning equilibrium: knowing the other players are using FCL, the rational choice is to use FCL. The reason why we did not compare FCL to more intelligent players was to illustrate its robustness to defectors, in contrast to less naive agents which are easier to cooperate with (like FFQ). The robustness of FFQ to selfish players could be compared with the robustness of FCL to selfish players: they would be similar, the difference being the fact that FCL can start cooperating if the opponent stops defecting, while FFQ only converges (by construction) to a defensive behaviour when facing a selfish player.

\subsection{Implementation details}
We implemented FCL using a state-dependent learning rate $\alpha_t = (\sum_{l<t}\delta\{s^l=s^t\})^{-1}$ that counts the number of state visits, and explorations are determined by a process $\phi(t) = \{X_t>\epsilon d^t\}$ where $X_t$ is a pseudo-random uniform sample between 0 and 1 with a communicated seed, $\epsilon$ a threshold and $d$ a decay parameter. For selfish $Q$-learning, we used a similar learning rate and exploration process, however with different seeds and decay parameters. The policy gradient was implemented with a tabular representation and Adam gradient descent with learning rate 0.1. Since matrix games are not sequential and since grid games were automatically reset after 30 steps, we could use a discount factor $\gamma=1$ to estimate value functions. In practice, we found that adding 1 to the minimal number of retaliation repeats given in Eq.~\ref{punish_repeats} significantly improves the robustness to selfish learners. In iterated matrix games, since they do not require large explorations, we used $\epsilon=0.5$ and $d=0.9$ for both selfish $Q$-learning and FCL. We used $\epsilon=1$ and $d=0.995$ in grid games.

\subsection{Results}

\begin{figure}[t]
    \centering
    \includegraphics[width=1\textwidth]{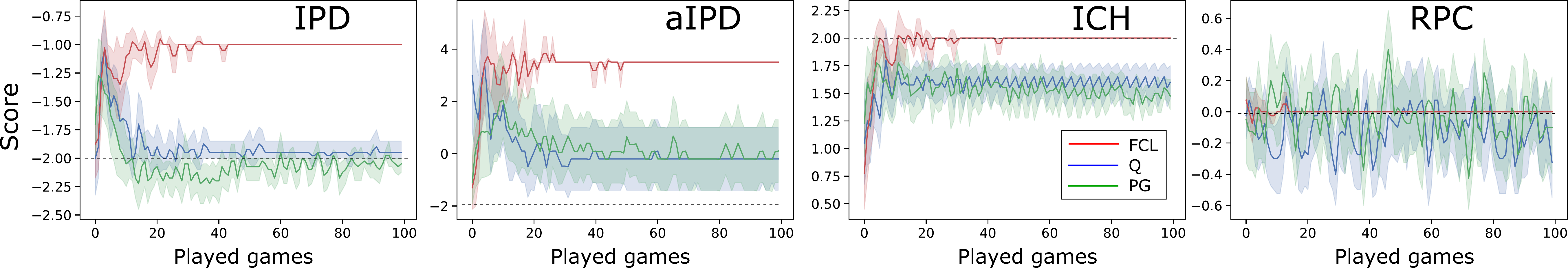}
    \caption{Matrix games. Average scores over 20 runs obtained by two standard RL algorithms and FCL, playing against FCL. In IPD and ICH, after some iterations selfish behaviors, as induced by Q-learning and PG, start being sub-optimal because of FCL retaliations and accumulate less return than a cooperative behaviors, as induced by FCL against itself. In RPS, FCL learns to play with a uniform distribution against selfish algorithms so their average score is null. Black dotted line represents the average score after convergence of two selfish agents playing against themselves (the \textit{minimax} solution).}
    \label{fig:matrix}
\end{figure}

\begin{figure}[t]
    \centering
    \includegraphics[width=\textwidth]{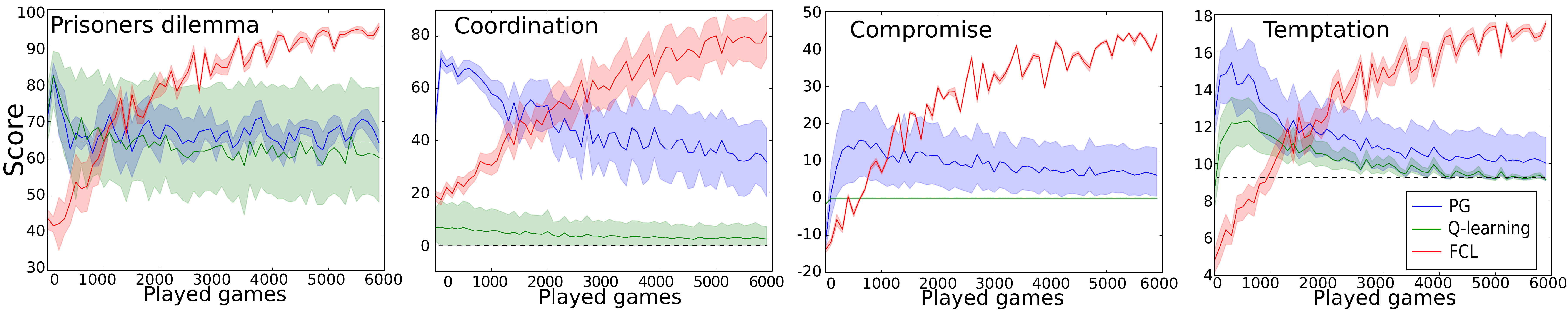}
    \caption{Grid games. Average scores over 20 runs obtained by two standard RL algorithms and FCL, playing against FCL. After some iterations, selfish behaviors, as induced by Q-learning and PG, start being sub-optimal because of FCL retaliations and accumulate less return than a cooperative behavior, as induced by FCL against itself. Black dotted line represents the average score after convergence of two selfish agents playing against themselves (the \textit{minimax} solution).}
    \label{fig:grid_games}
\end{figure}

In order to experimentally verify that FCL behaves as a learning equilibrium, we observe the score obtained by its opponent: if the opponent cooperates (for example if the opponent is itself an FCL), it should always obtain a better score than by following any other behavior (if the opponent follows a selfish policy gradient or a Q-learning algorithm). Therefore we compare the score of an FCL player with the scores of policy-gradient and Q-learning players, while playing with an FCL opponent. Figure~\ref{fig:matrix} displays our results with the three matrix games IPD, aIPD, ICH and RPS. Figure~\ref{fig:grid_games} displays our results on grid games. As expected, the score of selfish learners was never higher than the score of FCL, when the opponent is FCL. Except in RPS, defection conduced to less rewards than cooperation because of retaliations. In RPS, FCL found that the only way to retaliate was by infinitely playing randomly against selfish learners, resulting in an average of 0 reward for all players, equivalent to the reward for cooperation. This illustrates the fact that FCL is a learning equilibrium, since no tested algorithm performs better than FCL against FCL. Consequently, FCL was never exploited by selfish learners while being cooperative in self-play. 

\begin{wrapfigure}{r}{0.6\textwidth}
    \vspace{-27pt}
    \centering
    \includegraphics[width=0.4\textwidth]{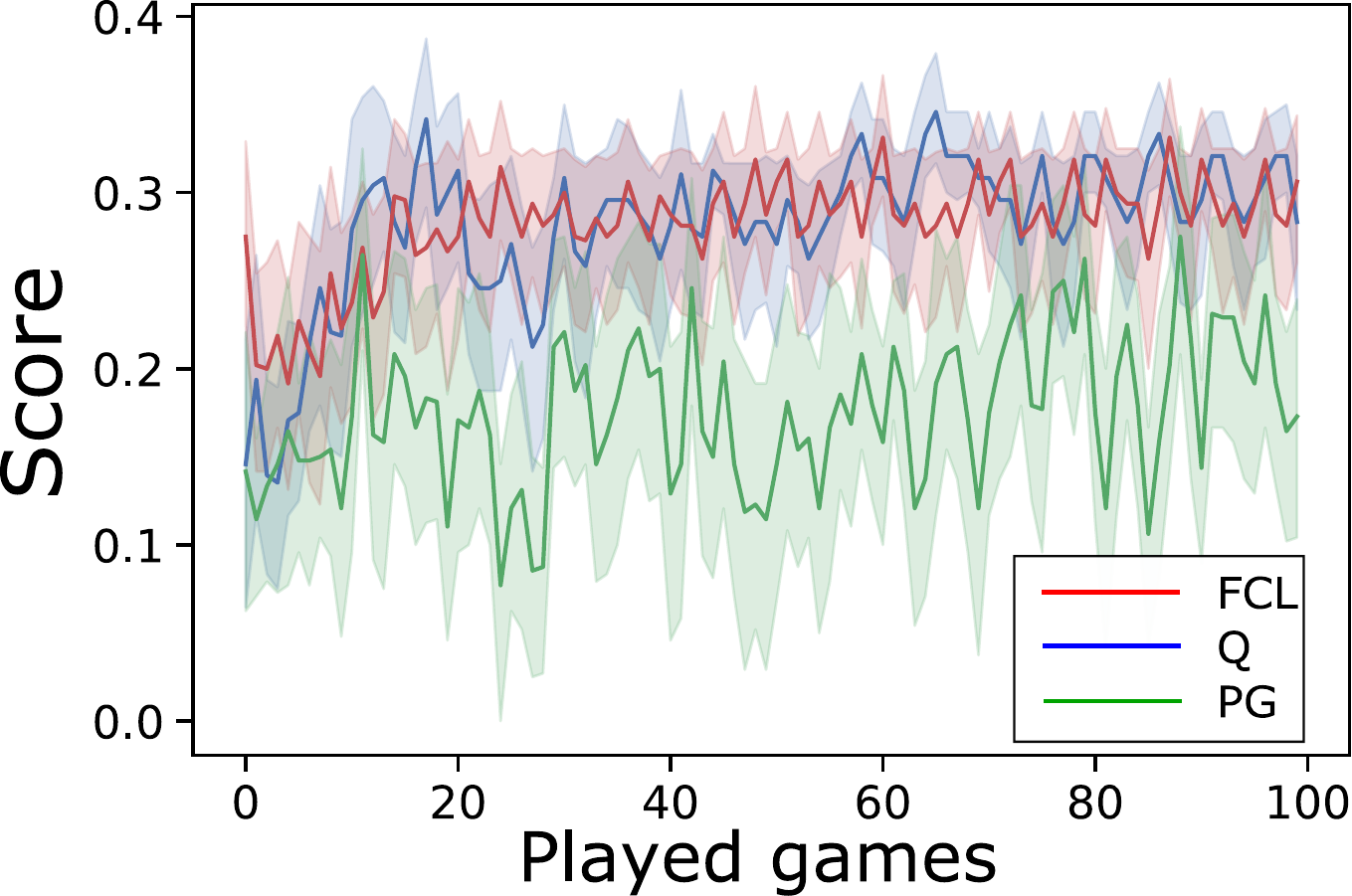}
    \caption{Grid games. Average scores over 20 runs obtained by two standard RL algorithms and FCL, playing against a team of 2 FCL. Q-learning players learned to cooperate with the team of FCL, leading to the same score than the one obtained by three FCL players ($1/3$ if all players chose to share).}
    \label{fig:cake}
    \vspace{-27pt}
\end{wrapfigure}

\subsection{More than 2 players}

We also investigated the robustness of FCL when more than 2 players are involved. What is expected is that the best a single agent can do when facing a team of FCL is to play as FCL. In order to verify this clam, we created a social dilemma game involving $N>2$ players. It is a tensor game where $N$ agents are given a cake, and may choose between sharing, robing or poisoning. If no player robs, each sharing player receives a reward 
$\frac{1- c}{n_s}$, where $n_s$ is the number of sharing agents and $c=-\frac{n_p}{(N-1}$ the eventual cost from poison (where $n_p$ is the number of poisoning agents). If $n_r$ agents decide to rob, one of them receives a reward $0.5 - c$ with probability $\frac{1}{n_r}$, all others receive nothing. Poisoning agents receive no reward. The cost from poison is designed to equal one if all but one agent poisoned the cake while the last one decided to rob it. If all agents cooperate, the best they can do is to share the cake, obtaining  a reward of $\frac{1}{N}$ (0.333 if $N=3$). But if one single agent starts to rob the cake (while all other are sharing) , it obtains a reward of 0.5. Figure~\ref{fig:cake} shows the results when a team of two FCL are facing a Q-learning, a PG or a third FCL. It appears that Q-learning players learned to cooperate with the team of FCL, leading to the same score than the one obtained by three FCL players. 

\section{Related work}

Learning cooperative behaviors in a multi-agent setting is a vast field of research, and various approaches depend on assumptions about the type of games, the type and number of agents, the type of cooperation and the initial knowledge. 

When the game's dynamics are initially known and in two-player settings, an egalitarian strategy can be obtained by mixing dynamic and linear programming. Therefore, a polynomial-time algorithm can be used to solve repeated matrix games~\citep{littman2005polynomial}, as well as repeated stochastic games~\citep{de2012polynomial}. Since an egalitarian strategy is always better than a \textit{minimax} strategy (the disagreement point)~\citep{osborne1994course}, a cooperative Nash equilibrium is immediately given. An alternative to our cooperate or retaliate architecture consists in choosing between maximizing oneself reward (being competitive) or maximizing a cooperative reward, for example by inferring opponents intentions~\citep{kleiman2016coordinate}. In contrast, FCL does not require the model in order to construct a foolproof cooperative behavior.

In games inducing social dilemmas and when the dynamics are accessible as an oracle, cooperative solutions can also be obtained by self-play and then applied to define a retaliating behavior forcing cooperation~\citep{lerer2017maintaining}, even when opponent actions are unknown, since in that case the reward function already brings sufficient information~\citep{peysakhovich2017consequentialist}. Here again, they use an offline procedure which does not apply to our purely online setting.

Closer to our setting, when the dynamics are unknown, online MARL can extract cooperative solutions in some non-cooperative games, and particularly in restricted resource appropriation~\citep{perolat2017multi}. Using alternative objectives based on all players reward functions and their propensity to cooperate or defect improves and generalizes the emergence of cooperation in non-cooperative games and limits the risk of being exploited by purely selfish agents~\citep{hughes2018inequity}. Like FFQ~\citep{littman2001friend}, the novelty of FCL is to disentangle the cooperative and the retaliating strategies so it can always switch from one behavior to the opposite without a forgetting and re-learning phase.

A similar approach, called Learning with Opponent Learning Awareness (LOLA), consists in modelling the strategies and the learning dynamics of opponents as part of the environment's dynamics and to derive the gradient of the average return's expectation~\citep{foerster2018learning}. If LOLA has no guaranties of convergence, a recent improvement of the gradient computation, which interpolates between first and second-order derivations, is proved to converge to a local optimums~\citep{letcher2018stable}. Although such agents are purely selfish, empirical results show that they are able to shape each others learning trajectories and to cooperate in prisoners dilemma. A limitation of this approach towards building learning equilibrium is the strong assumption regarding the opponents learning algorithms, supposed to perform policy gradient. Also, this approach differs to our goal since LOLA is selfish and aims at shaping an opponent's behavior (in 2-player settings) while FCL is cooperative but retaliates in response to selfish agents (in $N$-player settings).

Learning equilibrium solutions have been constructed for repeated matrix games~\citep{brafman2003efficient,ashlagi2012robust}. However, these solutions would not easily adapt to stochastic games, one main reason being the fact that exploration becomes infinite, while it only requires $A\times N$ steps in $N$-agent matrix games with $A$ different actions. Consequently, after a finished phase of exploration in matrix games, the deterministic payoff matrix is known and they can extract a Nash Equilibrium to exploit. Note that the restriction to symmetric games seems recurrent in learning equilibrium literature~\citep{brafman2005optimal,tennenholtz2009learning}. In repeated congestion games, it is even possible to construct a class of asymmetric games that does not admit any learning equilibrium, hence the importance of the symmetry assumption.

\section{Conclusion}

We introduced FCL, a model-free learning algorithm that, by construction, converges to an equilibrium strategy, cooperative with itself and retaliating when selfish algorithms are defecting. We proposed a definition for learning equilibrium, describing a class of learning algorithms such that the best way to play against it is to adopt the same behavior. We demonstrated that FCL is a learning equilibrium that forces a cooperative behavior, and we empirically verified this claim in various settings. Our approach could be improved by facilitating opponent's learning of the optimal cooperative response and by using faster learning approaches. It could also be adapted to larger dimensions such as continuous state spaces and partially observed settings with function approximation by replacing tabular $Q$-learning with deep $Q$-learning~\citep{mnih2015human}.

\appendix

\section{Sufficient number of retaliations.}\label{retaliations}
\begin{proof}
    Since $K_j \geq \frac{V^d_j - V^c_j}{V^c_j - V^r_j}$ and $V^c_j \geq V^r_j$, we write:
    \begin{equation}
        K_j (V^c_j - V^r_j) \geq V^d_j - V^c_j,
    \end{equation}
    which gives:
    \begin{equation}
        V^c_j \geq \frac{1}{K_j+1} (V^d_j + K_j V^r_j).
    \end{equation}
    On the left, this is the average return over stages of an always cooperating player, on the right this is the average return over stages of any deviating player. Therefore, for any $\{\pi^t_j\}_t\neq\{\boldsymbol{\pi}^t_\text{Folk}\}_t$:
    \begin{equation}
      \EE_{t\geq 0}\left[\R_j(\boldsymbol{\pi}^t_\text{Folk})\right] \geq \EE_{t\geq 0}\left[\R_j(\pi^t_j,\pi^t_{\text{Folk}_{-j}})\right].
    \end{equation}
\end{proof}

\section{Proof of Thm.~\ref{egalitarian}}\label{proof_egalitarian}

\begin{proof}
    Since $\sigma$ is N-cyclic, any player $i$ receives the same average return every N stage games:
    \begin{align}
        \EE_{t\geq 0}\left[\R_i(\pi^t)\right] &= \frac{1}{N}\sum_{t=1}^N \R_i(\pi^\Sigma_{\sigma^t(i)}, \pi^\Sigma_{\sigma^t(-i)})
        \\
        &\textit{ (the strategy is stationary)}
        \\
        &= \frac{1}{N}\sum_{t=1}^N \R_{\sigma^{t}(i)}(\pi^\Sigma_i, \pi^\Sigma_{-i}) 
        \\
        &\textit{ (the game is symmetric)}
        \\
        &= \frac{1}{N}\sum_{t=1}^N \R_t(\pi^\Sigma_i, \pi^\Sigma_{-i}) 
        \\
        &\textit{ (changing the order)}.
    \end{align}
    Consequently, $\pi_t$ maximizes the sum of returns at any $t$ and the average return of the strategy is the same for all players. Now, imagine there is a strategy $\{\hat{\pi}^t\}_t$ such that:
    \begin{equation}
        \min_i\EE_{t\geq 0}\left[\R_i(\hat{\pi}_i^t, \hat{\pi}_{-i}^t)\right]>\min_i\EE_{t\geq 0}\left[\R_i(\pi^t_i,\pi^t_{-i})\right].
    \end{equation}
    In that case, 
    \begin{align}
    \sum_i\EE_{t\geq 0}\left[\R_i(\hat{\pi}_i^t, \hat{\pi}_{-i}^t)\right] &>N\min_i\EE_{t\geq 0}\left[\R_i(\pi^t_i,\pi^t_{-i})\right]
    \\
    &=\sum_i\EE_{t\geq 0}\left[\R_i(\pi^t_i,\pi^t_{-i})\right],
    \end{align}
    which is in contradiction with the fact that $\pi_t$ maximizes the sum of returns at any $t$.
\end{proof}

\section{Proof of Thm.~\ref{fcl}}\label{proof_fcl}

\begin{proof}
    Let first show that FCL converges to cooperation in self-play (when all agents are playing FCL). Given the fact that $\phi$ is deterministic, there are no defections. Let's focus on $\{\hat{\pi}^t\}_t = \{\pi^t \vert \phi(t)=\textbf{True}\}$ the endless sub-process corresponding to exploration times. Let $\mathbb{A}=\A_1\times\dots\times\A_N$. Clearly, for all $t$, $s\in\s$ and $a\in\mathbb{A}$, $\hat{\pi}^t(a\vert s)>0$. Consequently, the convergence of $Q$-functions is given by the convergence of the classic $Q$-learning using $\pi(a^t=a\vert s^t=s)=\hat{\pi}^t(a\vert s)$~\cite{melo2001convergence}. Let ${Q^c}^*$ and ${Q^r_j}^*$ be the corresponding points of convergence. By construction:
    \begin{itemize}
        \item $\boldsymbol{\pi^c}$, maximizing ${Q^c}^*$, maximizes the sum of players returns.
        \item $\boldsymbol{\pi^*} = (\pi^c_{\sigma^t(1)}\dots \pi^c_{\sigma^t(N)})$ maximizes the min of players returns.
        \item $\boldsymbol{\pi^r}(j)$, minimizing ${Q^r_j}^*$, retaliates when player $j$ deviates from $\boldsymbol{\pi^*}$.
    \end{itemize}
    FCL decision rule at line 7 in Alg.\ref{algo:FCL} corresponds to playing according to $\boldsymbol{\pi^*}$ when $Q^c$ is close enough to ${Q^c}^*$ (when the difference between the max value and the second-max value is larger than twice an update size). Similarly, decision rule at line 11 corresponds to playing according to $\boldsymbol{\pi^r}(j)$ when $Q^r_j$ is close enough to ${Q^r_j}^*$. Let $T_Q$ be the smallest time after which both $Q^c$ and $Q^r_j$ are close enough to ${Q^c}^*$ and ${Q^r_j}^*$ so lines 7 and 11 correspond to playing according to $\boldsymbol{\pi^*}$ and $\boldsymbol{\pi^r}(j)$. If explorations are stopped, FCL behaves as a cooperative equilibrium.

    Now, let $A_j = \{\pi_j^t\}_t$ be a learning algorithm different than FCL and played by an agent $j$ while all other players follow FCL ($A_{-j} = \{\pi_{-j}^{t}\}_t$). 
    We will distinguish two situations:
    \begin{align}
        \text{(a) } \forall s\in\s\, , \,\forall a_j\in\A_j\, , &\,\forall a_{-j}\in\A_{-j}\, , \,\forall T\, , \, \exists t>T, 
        \pi_j^t(a_j\vert s)\times \pi_{-j}^t(a_{-j}\vert s)>0,\\
        \text{(b) } \exists s\in\s\, , \,\exists a_j\in\A_j\, , &\,\forall a_{-j}\in\A_{-j}\, , \,\exists T\, , \, \forall t>T, 
        \pi_j^t(a_j\vert s)\times \pi_{-j}^t(a_{-j}\vert s)=0.
    \end{align}
    In situation (a), all possible transitions are explored: the conditions for the convergence of FCL agents are met and they converge to a cooperative equilibrium, retaliating if the non-FCL learner is not cooperative. However, we must make sure that FCL's explorations are not spoiling the efficiency of retaliations:
  Let $\R_1 = \EE_{t>T_\epsilon}[\R_j(A_j, A^*_{-j})]$ be a deviating player's average return when FCL players are greedy (with probability $1-\epsilon$), $R_2$ be its average return when other agents are exploring (with probability $\epsilon$, and $\R^*_1$ and $\R^*_2$ be the respective returns when no player deviates). Because the game is symetric, we know that $\R_1 \leq \R^*_1$. If $\R_2\leq \R^*_2$, then the average return of a deviating player is always smaller than if it does not deviate. Otherwise, we have $\R_2 + \R_1^* > \R_2^* + \R_1$ and by taking:
    \begin{equation}
        \epsilon = \frac{\R_1^* - \R_1}{\R_2 + \R_1^* - (\R_2^* + \R_1)}
    \end{equation}
    we obtain, for all $t>T_\epsilon$:
    \begin{equation}
        (1-\epsilon)\R_1 + \epsilon \R_2 \leq (1-\epsilon)\R_1^* + \epsilon \R_2^*.
    \end{equation}
    Consequently:

    \begin{align}
        \EE_{t>T_\epsilon}\big[\R_j\big(A_j(t), & A_{-j}^*(t)\big)\big] \leq \EE_{t>T_\epsilon}\big[\R_j\big(A_j^*(t), A_{-j}^*(t)\big)\big].
    \end{align}
    
    In situation (b), there is still a subset of state-action couples $\Omega_\infty$ that will be explored an infinite number of times. If all other players restrict their states and actions to the same subset (using $\pi_i(a\vert s)>0 \Leftrightarrow (a,s)\in\Omega_\infty$) the induced sub-game is still symmetric and player $j$ is exploring the whole sub-game an infinite number of times. 
    Consequently, FCL can at least learn a cooperative equilibrium $\{\boldsymbol{\hat{\pi}}^t_\text{Folk}\}_t$ based on a retaliation strategy $\boldsymbol{\pi}^{r,j}_{\Omega_\infty}$ and a cooperative strategy $\boldsymbol{\pi}^*_{\Omega_\infty}$ defined on $\Omega_\infty$ such that:
    \begin{align}
        \forall \{\pi_j^t\}_t \neq \{\boldsymbol{\hat{\pi}}^t_{\text{Folk}_t}\},\;
        \EE_{t\geq 0}[\R_j(\boldsymbol{\hat{\pi}}^t_\text{Folk})] &= \EE_{t\geq 0}[\R_j(\boldsymbol{\pi}^*_{\Omega_\infty})]
        \geq \EE_{t\geq 0}[\R_j(\pi^t_j,\boldsymbol{\hat{\pi}}^t_{\text{Folk}_{-j}})].
    \end{align}
    Since $\boldsymbol{\pi}^*_{\Omega_\infty}$ is necessarily sub-optimal to cooperate in the whole game, we have:
    \begin{align}
        \forall \{\pi_j^t\}_t \neq \{\boldsymbol{\hat{\pi}}^t_{\text{Folk}_t}\},\;
        \EE_{t\geq 0}[\R_j(\boldsymbol{\pi}^*)] &\geq \EE_{t\geq 0}[\R_j(\boldsymbol{\pi}^*_{\Omega_\infty})]
        \geq \EE_{t\geq 0}[\R_j(\pi^t_j,\boldsymbol{\hat{\pi}}^t_{\text{Folk}_{-j}})].
    \end{align}
    As a consequence, players can still retaliate and we can use the exact same argument than in (a) to obtain the desired statement.
\end{proof}

\bibliography{biblio}

\end{document}